\documentstyle[preprint,eqsecnum,aps]{revtex}

\begin{document}

\draft 
\preprint{}
\topskip 2cm

\begin{center}
{\bf DIMENSIONALLY CONTINUED OPPENHEIMER-SNYDER\\ 
GRAVITATIONAL COLLAPSE II:  SOLUTIONS IN ODD DIMENSIONS} \\
\vskip 2cm
{\bf Anderson Ilha} \\
\vskip 0.3cm
{\scriptsize  Departamento de Astrof\'{\i}sica,
	      Observat\' orio Nacional-CNPq,} \\
{\scriptsize  Rua General Jos\'e Cristino 77,
	      20921-400 Rio de Janeiro, Brazil.} \\
\vskip 0.6cm
{\bf Antares Kleber} \\
\vskip 0.3cm
{\scriptsize  Departamento de Astrof\'{\i}sica,
	      Observat\' orio Nacional-CNPq,} \\
{\scriptsize  Rua General Jos\'e Cristino 77,
	      20921-400 Rio de Janeiro, Brazil.} \\
\vskip 0.6cm
{\bf Jos\'e P. S. Lemos} \\
\vskip 0.3cm
{\scriptsize  Departamento de Astrof\'{\i}sica,
	      Observat\' orio Nacional-CNPq,} \\
{\scriptsize  Rua General Jos\'e Cristino 77,
	      20921-400 Rio de Janeiro, Brazil.} \\
{\scriptsize  \&} \\
{\scriptsize  Departamento de F\'{\i}sica,
	      Instituto Superior T\'ecnico,} \\
{\scriptsize  Av. Rovisco Pais 1, 1096 Lisboa, Portugal.} \\

\end{center}

\title{Dimensionally Continued Oppenheimer-Snyder\\ 
Gravitational Collapse: Solutions in Odd Dimensions}
\author{Anderson Ilha}
\address{Departamento de Astrof\'{\i}sica,
	      Observat\' orio Nacional-CNPq, \\
              Rua General Jos\'e Cristino 77,
	      20921-400 Rio de Janeiro, Brazil,}

\author{Jos\'e P. S. Lemos}
\address{Departamento de Astrof\'{\i}sica,
	      Observat\' orio Nacional-CNPq, \\
              Rua General Jos\'e Cristino 77,
	      20921-400 Rio de Janeiro, Brazil,  \\  \& \\
              Departamento de F\'{\i}sica,
	      Instituto Superior T\'ecnico, \\
              Av. Rovisco Pais 1, 1096 Lisboa, Portugal,}

\vfill\eject

\begin{abstract}
The Lovelock gravity extends the theory of general relativity to higher
dimensions in such a way that the field equations remain of second
order. The theory has many constant coefficients with no a priori
meaning. Nevertheless it is possible to reduce them to two, the
cosmological constant and Newton's constant. In this process one
separates theories in even dimensions from theories in odd dimensions.
In a previous work gravitational collapse in even dimensions was
analysed.  In this work attention is given to odd dimensions. It is
found that black holes also emerge as the final state of gravitational
collapse of a regular dust fluid.
 
\end{abstract}
\bigskip

\widetext

\section{Introduction}
\label{sec:level1}

A generalization of Einstein gravity to other dimensions while keeping
the same degrees of freedom (the field equations for the metric remain
of second order) is given by the Lovelock action  \cite{Love}.
The theory can also be considered as an extension of 
Eintein-Hilbert action (see e.g. \cite{TZ}), in which new terms make their 
appearance by taking into the action the Euler densities of the spaces 
with dimensions lower than the space in consideration.

In a previous work \cite{aisjps01} we have studied gravitational
collapse in Lovelock gravity for a spacetime with even dimensions,
thus extending the Oppenheimer-Snyder collapsing model.
Following the work of \cite{TZ,BTZ}, the reason for separating even
from odd dimensions in the Lovelock theory comes naturally in a $\cal
D-$dimensional spacetime when one considers embedding the Lorentz group
$SO({\cal D}-1,1)$ into de anti-de Sitter group $SO({\cal D}-1,2)$. The
Lovelock theory then branches into two distinct classes,  with
Lagrangians for even dimensions and Lagrangians for odd dimensions. One
also finds in this way that the number of constants, which proliferates
when one goes to higher and higher dimensions, reduces drastically to
two, the cosmological constant $\Lambda$ and Newton's constant $G$.

In this work we study gravitational collapse in odd-dimensional
spacetimes and show that black holes form from regular initial data
consisting of a dust fluid. We follow closely the nomenclature and the
division of sections made in  \cite{aisjps01}.  In section II the
Lovelock gravity for restricted coefficients in odd-dimensional
spacetimes is presented.  In section III we display the static
solutions in odd dimensions found in \cite{BTZ}. In section IV we find
some cosmological or interior matter solutions for perfect fluids. In
section V we match the solutions found in section IV to the solutions
of section III. In section VI we show that black holes can form through
gravitational collapse in Lovelock odd-dimensional gravity. Section VI
comments on the formation of naked singularities and section VII
presents some conclusions.  In the paper we usually do $G=c=1$.

\section{The Lovelock theory}

The most general action in ${\cal D} \geq 3$ spacetime dimensions that
yields    the same degrees of freedom of Einstein's theory is the so
called Lovelock action, given by \cite{Love,TZ}
\begin{equation}
S = \int\, {\cal L}_{\cal D} =
\kappa\,\sum_{p=0}^{\left[({\cal D}-1)/2\right]}\,
\alpha_{p}\,\int_{M}\,
\epsilon_{a_{1} \cdots
a_{{\cal D}}}\,R^{a_{1}a_{2}}\,\wedge\,\cdots\,
\wedge\,R^{a_{2p-1}a_{2p}}\,
\wedge\,e^{a_{2p+1}}\,\wedge\,\cdots\,\wedge\,
e^{a_{{\cal D}}} + S_{m}\, ,
\label{2.1}
\end{equation}
where $R^{ab}={d}\omega^{ab}+
\omega^{a}_{\,\,c}\,\wedge\,\omega^{cb}$ is the curvature two-form,
$e^{a}$ is the local frame one-form,  and $\omega^{ab}$ is the spin
connection, with $a_{i} = {0, 1, \dots, {\cal D} -1}$.  The symbol $[
]$ over the summation symbol means one should take the integer part of
$({\cal D}-1)/2$. $S_{m}$ is a phenomenological action wich describes
the macroscopic matter sources.

In general, the constant coefficients $\alpha_{p}$ are arbitrary.
However, it is shown in \cite{BTZ} that taking certain special choices 
one is able to get simple meaningful solutions. Following \cite{BTZ} one
first considers embedding the Lorentz group $SO({\cal D}-1,1)$ into de
anti-de Sitter group $SO({\cal D}-1,2)$, and then separates into two
distinct classes of Lagrangians: Lagrangians for even dimensions and 
Lagrangians for odd dimensions.

For odd dimensions, ${\cal D} = 2n-1$, one can find a construction
similar to the Chern-Simons action construction in three dimensions.
One starts with the Euler density in one dimension above ${\cal D}$,
\begin{equation}
E_{2n}=\kappa\,\epsilon_{A_{1} \cdots A_{2n}}\,\hat{R}^{A_{1}A_{2}}\,
\wedge\,\cdots\,\wedge\,\hat{R}^{A_{2n-1}A_{2n}}\, ,
\label{2.2}
\end{equation}
with $A_{1}, A_{2} = 0, 1, \dots \, 2n-1$ being the anti-de Sitter
indices. $\hat{R}^{AB}$ is the anti-de Sitter curvature two-form
constructed with the SO(${\cal D}$-1,2) connection $W^{AB}$.
Equation (\ref{2.2}) is a local exact form, and can be written as an
exterior derivative of a Lagrangian in $2n-1$ dimensions, i.e.,
$E_{2n} = {d}{\cal L}_{2n - 1}$, see \cite{BTZ}. Decomposing the
connection $W^{AB}$ into the connection under ${\cal D}$ rotations
$w^{ab}$ and inner translations $e^{a}$, one finds the anti-de Sitter
curvature $\hat{R}$ in terms of the Lorentz curvature $R$:
\begin{equation}
\hat{R}^{ab}=R^{ab} + \frac{1}{l^{2}}\,e^{a}\,\wedge\,e^{b}\, ,
\label{2.3}
\end{equation}
where $l$ is a scale factor which is to be related to the cosmological
constant $l^{2}=-1/\Lambda$. Using Eq. (\ref{2.3}) one finds that the
Lagrangian in Eq. (\ref{2.2}) can be put in the form
\begin{equation}
{\cal L}_{2n-1} = \kappa\,\sum_{p=0}^{n-1}\,\alpha_{p}\,
\epsilon_{a_{1}\cdots a_{{\cal D}}}\,R^{a_{1}a_{2}}\,\wedge\,
\cdots\,\wedge\,R^{a_{2p-1}a_{2p}}\,\wedge\,e^{a_{2p+1}}\,\wedge\,
\cdots\,\wedge\,e^{a_{{\cal D}}}\,
\label{2.4}
\end{equation}
where the coeficcients are given by
\begin{equation}
\alpha_{p}=\frac{1}{{\cal D}-2p}\,\left(
                                  \begin{array}{c}
                                  n - 1  \\ p 
                                  \end{array}
                                  \right)\,
l^{-{\cal D}+2p}\,,
\label{2.5}
\end{equation}
and for convenience one can choose $\kappa$ as
\begin{equation}
\kappa = \frac{{\cal D}-2}{16\,\pi\,G\,n}\,l^{{\cal D}-2}\, .
\label{2.6}
\end{equation}

Given the action (\ref{2.1}), the field equations are obtained by the
variation with respect to the one-forms $e^{a}$.  Under the assumption
of zero torsion, the variation with respect to the spin connection
$\omega^{ab}$ vanishes identically. Although the equations have powers
in the curvatures, they remain by construction  second order in the
metric. The field equations are given by
\begin{equation}
-\kappa\,\sum_{p=0}^{\left[({\cal D}-1)/2\right]}\,
\alpha_{p}\,({\cal D} - 2p)\,\epsilon_{a_{1} ..
a_{{\cal D}}}\,R^{a_{1}a_{2}}\,\wedge\,..\,\wedge\,
R^{a_{2p-1}a_{2p}}\,
\wedge\,e^{a_{2p+1}}\,\wedge\,..\,\wedge\,
e^{a_{{\cal D}-1}} = Q_{a_{{\cal D}}}\, ,
\label{2.7}
\end{equation}
where $Q_{a_{{\cal D}}}$ is a $({\cal D} -1)$-form
associated with the energy momentum tensor $T^{a}_{b}$ through the following
expression
\begin{equation}
Q_{i} = \frac{1}{({\cal D} -1)!}\,T^{a_{1}}_{i}\,
\epsilon_{a_{1} \cdots a_{{\cal D}}}\,e^{a_{2}}\,\wedge\,\cdots\,\wedge\,
e^{a_{{\cal D}}}\, .
\label{2.8}
\end{equation}

\section{Exterior Vacuum Solutions}

In the vacuum  all components of the energy-momentum tensor vanish, so that 
the field equations (\ref{2.7}) are given by
\begin{equation}
-\kappa\,\sum_{p=0}\,\alpha_{p}\,({\cal D} - 2p)\,
\epsilon_{a_{1} \cdots
a_{{\cal D}}}\,R^{a_{1}a_{2}}\,\wedge\,\cdots\,\wedge\,
R^{a_{2p-1}a_{2p}}\,
\wedge\,e^{a_{2p+1}}\,\wedge\,\cdots\,\wedge\,
e^{a_{{\cal D}-1}} =  0\, .
\label{3.1}
\end{equation}
Inserting the coeficients  $\alpha_{p}$ and the constant $\kappa$ given
in (\ref{2.5}) and  (\ref{2.6}) in equation (\ref{3.1}), one gets 
for odd dimensions (${\cal D}=2n-1$), 
\begin{equation}
(R^{a_1 a_2} + l^{-2}\, e^{a_1}\, \wedge \,
e^{a_2})\, \wedge\, \cdots \, \wedge\,
(R^{a_{2n-3} a_{2n-2}} + l^{-2}\, e^{a_{2n-3}}\,
\wedge\, e^{a_{2n-2}})\,
\epsilon_{a_1 a_2 \cdots a_{2n-1}} = 0.
	\label{3.2}
\end{equation}
We consider now a static, spherical symmetric spacetime. 
One can write the metric in the following form, 
\begin{equation}
{d}s^2_{+}=-g^2(r_{+})\, 
{d}t_{+}^2 + g^{-2}(r_{+})\,{d}r_{+}^2 + 
r_{+}^2\,{d}\Omega_{{\cal D}-2}^2\, ,
\label{3.3}
\end{equation}
where $t$ and $r$ are the time and radial coordinates and
${d}\Omega_{{\cal D}-2}^2$ is the arc-element of a unit (${\cal
D}-2$)-sphere.  The subscript + reminds that (\ref{3.3}) is to be
viewed  as an exterior solution.  With  metric (\ref{3.3}) and
equations (\ref{3.1}) and (\ref{3.2}), Ba\~nados, Teitelboim and
Zanelli found the following exact solution for ${\cal D}=2n-1$
\cite{BTZ},
\begin{equation}
{d}s^2_{+} = -\left[1 - (M + 1)^{2/({\cal D}-1)} +
 (r_{+}/l)^2\right]\,{d}t_{+}^2 +
\frac{{d}r_{+}^2}{1 - (M + 1)^{2/({\cal D}-1)} + (r_{+}/l)^2} + r_{+}^2\,
 {d}\Omega^2_{{\cal D}-2}. 
\label{3.4}
\end{equation}
These solutions describe black holes. We will show that they also
represent the exterior vacuum solution to a collapsing (or expanding) 
dust cloud in Lovelock's odd-dimensional theory, as in the 
even-dimensional case \cite{aisjps01}.

\section{Interior Matter Solutions}
	
The interior spacetime is modeled by a homogeneous 
collapsing (or expanding) dust cloud, whose metric is described by the
Friedmann-Robertson-Walker in ${\cal D}$ dimensions
\begin{equation}
ds^{2} = -dt^{2} + a^{2}\left(t\right)\,
 \left[\frac{dr^{2}}{1 - k\,r^{2}} + 
r^{2}\,d\Omega^{2}_{{\cal D}-2} \right]\, .
	\label{4.1}
\end{equation}
The coordinates $t$ and $r$ are comoving coordinates 
(we omit throughout the subscript $-$ to indicate an interior solution). 
Note that that $k$ has dimension of
$1/[\mbox{length}]^{2}$.
The energy-momentum tensor for a perfect fluid is given by
\begin{equation}
T_{\alpha\beta} = (\rho+p)\,u_{\alpha}\,u_{\beta} + p g_{\alpha\beta}\, ,
\label{4.2}
\end{equation}
where $\rho$ is the energy-density, $p$ the pressure, 
and $u^{\alpha}$ is the ${\cal D}$-velocity of the  fluid. 
From (\ref{4.1})-(\ref{4.2}) and 
Lovelock equations (\ref{2.7}) we obtain
\begin{equation}
-B\,\frac{d}{dt}\left(\frac{\dot{a}}{a}\right) 
+ \frac{k}{a^2} = \rho + p
\label{4.3}
\end{equation}
\begin{equation}
({\cal D} - 1)\, B\,\left(\frac{\dot{a}}{a}\right)\left[-\frac{k}{a^2} +
\frac{d}{dt}\left(\frac{\dot{a}}{a}\right)\right] = 
\dot{\rho}
\label{4.4}
\end{equation}
where
\begin{equation}
B \equiv ({\cal D}-2)!\,\sum_{p}\,\alpha_{p}\,2p\,({\cal D}-2p)\,
\left(\frac{\dot{a}^{2}+k}{a^2}\right)^{p-1}\, .
\label{4.5}
\end{equation}
where the coefficients $\alpha_{p}$ are given in (\ref{2.5}), and
$\kappa$ is given in (\ref{2.6}).
Equations (\ref{4.3})-(\ref{4.4}) have a first integral given by
\begin{equation}
\dot{a}^2 = -k - \left(\frac{a}{l}\right)^2 + \left(\frac{a_{0}}{l}\right)^2 
\left[ \frac{8\,\pi\,l^2\,
\rho_{0}}{({\cal D}-2)!\,({\cal D}-2)}\right]^{2/({\cal D}-1)}\, , 
\label{4.6}
\end{equation} 
where $\rho_0$ and $a_0$ are constants. 
Equations (\ref{4.3})-(\ref{4.4}) have also a second integral, i.e., 
the solution of the Eq. (\ref{4.6}) is given by (see also \cite{Li})
\begin{equation}
a(t/l) = \frac{l}{r_{\Sigma}}\,\sqrt{
\left\{\left(\frac{1}{l}\right)^{{\cal D}-3}\,\left[ 
\frac{8\,\pi\,\rho_{0}
\,(a_{0}\,r_{\Sigma})^{{\cal D}-1}}{({\cal D}-2)!\,({\cal D}-2)}\right]
 \right\}^{2/({\cal D}-1)} - 
k\,r_{\Sigma}^{2}}\,\sin (b + t/l)\, ,
\label{4.7}
\end{equation}
where $b$ is an arbitray phase which will be neglected henceforward.

The Ricci quadratic scalar and the Kretschmann scalar are given by
\begin{eqnarray}
&&R_{ab}\,R^{ab} = -\left({\cal D}-1\right)^{2}\,\left(\frac{\ddot{a}}{a}
\right)^{2}+\left({\cal D}-1\right)\,\left[\frac{\ddot{a}}{a} +
\left({\cal D}-2\right)\,\frac{\dot{a}^{2}+k}{a^{2}}\right]^{2}\, ,
\label{4.8} \\
&&R_{abcd}\,R^{abcd} = \left({\cal D}-1\right)\left[\left(\frac{
\ddot{a}}{a}\right)^{2} + \left(\frac{\dot{a}^{2} +k}{a^{2}}\right)^{2}
\right]\, ,
\label{4.9}
\end{eqnarray}
respectively.

We now assume a dust fluid, $p=0$. For such an equation of 
state we have 
\begin{equation}
\rho = \rho_{0}\,\left(\frac{a_{0}}{a}\right)^{{\cal D}-1}\, , 
\label{4.10}
\end{equation}
where $\rho_0$ and $a_0$ are the constants defined above.

Inserting Eq. (\ref{4.7}) in Eq. (\ref{4.10}), we obtain the evolution
of the density in the dust model:
\begin{equation}
\rho(t/l) = \rho_{0}\,\left[\frac{a_{0}\,r_{\Sigma}/l}
{\sqrt{
\left\{\left(\frac{1}{l}\right)^{{\cal D}-3}\,\left[ 
\frac{8\,\pi\,\rho_{0}
\,(a_{0}\,r_{\Sigma})^{{\cal D}-1}}{({\cal D}-2)!\,({\cal D}-2)}\right]
\right\}^{2/({\cal D}-1)} - k\,r_{\Sigma}^{2}}}\right]^{{\cal D}-1}\,
\sin^{-({\cal D}-1)}(t/l)\, .
\label{4.11}
\end{equation}
We see that the density (\ref{4.11}) and the curvature scalars
(\ref{4.8})-(\ref{4.9}) diverge
at $t/l = \pi$ which represents the formation of a singularity.

\section{Junction Conditions}

Now we match the exterior and interior spacetimes found in sections III 
and IV, respectively, across an interface of separation $\Sigma$. The 
junctions conditions are \cite{Israel}  
\begin{eqnarray}
\left. ds^{2}_{+}\right]_{\Sigma} &=& \left. ds^{2}_{-}\right]_{\Sigma} \label{5.1} \\ 
\left. K_{\alpha\beta}^{+}\right]_{\Sigma} &=& \left. 
K_{\alpha\beta}^{-}\right]_{\Sigma} \label{5.2}
\end{eqnarray}
where $K_{\alpha\beta}$ is the extrinsic curvature, 
\begin{equation}
K_{\alpha\beta}^{\pm}= -n_{\epsilon}^{\pm}\,\frac{\partial^{2}
x_{\pm}^{\epsilon}}
{\partial \xi^{\alpha}\partial \xi^{\beta}} - n_{\epsilon}^{\pm}\,
\Gamma_{\gamma\delta}^{\epsilon}\,
\frac{\partial x_{\pm}^{\gamma}}{\partial \xi^{\alpha}}\,
\frac{\partial x_{\pm}^{\delta}}{\partial \xi^{\beta}}
\label{5.3}
\end{equation}
and $n_{\epsilon}^{\pm}$ are the components of the unit normal vector to
$\Sigma$ in the coordinates $x_{\pm}$, and $\xi$ represents 
the intrinsic coordinates in $\Sigma$.  
The subscripts $\pm$
represent the quantities taken in the exterior and interior
spacetimes.  Both the metrics and the extrinsic curvatures in
(\ref{5.1})-(\ref{5.2}) are evaluated at $\Sigma$.  The metric
intrinsic to $\Sigma$  is written as 
\begin{equation}
ds^2_{\Sigma} = -d\tau^2 + R^2(\tau)\,
d\Omega^2_{{\cal D}-2}\, .
\label{5.4}
\end{equation}
Where $\tau$ is the proper time on $\Sigma$ and  $d\Omega^2_{{\cal D}-2}$
denotes the line element on a ${\cal D}-2$ dimensional sphere. 

Using the  junction condition (\ref{5.1}), metric (\ref{5.4}) 
and the exterior metric  (\ref{3.4}) we obtain 
\begin{equation}
\left. r_{+}\right]_{\Sigma} = R\left(\tau \right) \, ,
\label{5.5}
\end{equation}
and 
\begin{equation}
\left[1 - (M+1)^{2/({\cal D}-1)} + (r_{+}/l)^2\right]\,
\dot{t}_{+}^{2} -  \left[1 - (M+1)^{2/({\cal D}-1)} + 
(r_{+}/l)^2\right]^{-1}\, \dot{r}_{+}^{2} = 1\, \, ,
\label{5.6}
\end{equation}
where $\cdot\equiv d/d\tau$, and both equations are evaluated at
$\Sigma$. From now on, we will usually omit the subscript $\Sigma$ to 
denote evaluation at the interface.
Using (\ref{5.5}) in (\ref{5.6}) we find
\begin{equation}
\frac{dt_{+}}{d\tau} = \frac{\sqrt{\left[1 -
(M+1)^{2/({\cal D}-1)} + 
(R/l)^2\right] + \dot{R}^{2}}}{\left[1 - 
(M+1)^{2/({\cal D}-1)} + 
(R/l)^2\right]}\, .
\label{5.7}
\end{equation}

The unit normal to $\Sigma$ in the exterior spacetime is 
\begin{equation}
n_{\epsilon}^{+} = \left(-\frac{{d}r_{+}}{{d}\tau}, 
\frac{{d}t_{+}}{{d}\tau}, 0, \cdots, 0\right).  
\label{5.8}
\end{equation}
From (\ref{5.3}) we then get
\begin{equation}
K_{\theta\theta}^{+} = R\,\sqrt{\left[1 - \left(M+1
\right)^{2/({\cal D}-1} + 
\left(\frac{R}{l}\right)^2\right] + \dot{R}^{2}}\, .
\label{5.9}
\end{equation}
In what follows the other components of $K_{ab}^{+}$ are not needed. 

The unit normal to $\Sigma$ in the interior spacetime is 
\begin{equation}
n_{\epsilon}^{-} = \left(0, \frac{a}{\sqrt{1-k\,r^2}}, 0, \cdots, 0\right)
\label{5.10}
\end{equation}
and from  (\ref{5.3}) we have 
\begin{equation}
K_{\theta\theta}^{-} = 
R(\tau) \, \sqrt{1 - k\,r_{\Sigma}^2}\, .
\label{5.11}
\end{equation}
Using the  junction condition (\ref{5.1}) for the interior spacetime 
yields $ar_{\Sigma}=R(\tau)$.
From the condition $K_{\theta\theta}^{+} = K_{\theta\theta}^{-}$,  
(\ref{5.9}) and (\ref{5.11}) we obtain
\begin{equation}
\dot{R}^2 + \left(\frac{R}{l}\right)^2 + k\,
r_{\Sigma}^2 = \left(M+1
\right)^{2/({\cal D} - 1)}\, ,
	\label{5.12}
\end{equation}
Multiplying equation (\ref{4.6}) by $r_{\Sigma}^{2}$ 
we get
\begin{equation}
\dot{R}^2 + \left(\frac{R}{l}\right)^2 + k\,r_{\Sigma}^2 = 
\left(\frac{R_{0}}{l}\right)^2 \,
 \left[ \frac{8\,\pi\,l^2\,
\rho_{0}}{({\cal D}-2)!\,({\cal D}-2)}
\right]^{2/({\cal D} -1)}\, . 
\label{5.13}
\end{equation}
Comparing equation (\ref{5.12}) and (\ref{5.13}) we have 
\begin{equation}
M = \left(\frac{1}{l}\right)^{{\cal D}-3}\,\left[ 
\frac{8\,\pi\,\rho_{0}
\,(a_{0}\,r_{\Sigma})^{{\cal D}-1}}{({\cal D}-2)!\,
({\cal D}-2)}\right] - 1\,, 
\label{5.14}
\end{equation}
which is the mass of the cloud expressed in terms of the constants 
given in the problem. 

\section{Black Hole Formation}

In order to study black hole formation in this theory we work with the 
solution found in (\ref{4.7}). The interior and exterior metrics are
given in (\ref{4.1}) and in (\ref{3.4}) respectively, and as we have
shown in section V, it is possible to make a smooth junction between both 
spacetimes. To be complete we treat the cases  ${\cal D}\geq3$. 
The case ${\cal D}=3$ reduces to the collapse studied in \cite{MannRoss}. 

For convenience
we rewrite Eqs. (\ref{4.6})-(\ref{4.9}) and (\ref{4.11}) in terms
of the mass M. We have thus
\begin{equation}
a(t/l) = \frac{l}{r_{\Sigma}}\,\sqrt{\left(M+1\right)^{2/
({\cal D}-1)}-k\,r_{\Sigma}^{2}}\,\sin(t/l)\, ,
\label{6.1}
\end{equation}
for the scale factor,
\begin{equation}
\rho(t/l) = \rho_{0}\,\left[\frac{a_{0}\,r_{\Sigma}/l}
{\sqrt{(M+1)^{2/({\cal D}-1)}-k\,r_{\Sigma}^{2}}}\right]^{{\cal D}-1}\,
\sin^{-({\cal D}-1)}(t/l)\, ,
\label{6.2}
\end{equation}
for the density and
\begin{eqnarray}
R_{ab}\,R^{ab}=&&-\frac{\left({\cal D}-1\right)^{2}}{l^{4}} +
\frac{\left({\cal D}-1\right)}{l^{4}}\,\left\{-1+\right. \nonumber \\ 
&&+\left.\left({\cal D}-2\right)\,\frac{\left[\left(M+1\right)^{2/({\cal
D}-1)}-k\,r_{\Sigma}^{2}\right]\,\cos^{2}(t/l) + k\,r_{\Sigma}^{2}}
{\left[\left(M+1\right)^{2/({\cal D}-1)}-k\,r_{\Sigma}^{2}\right]\,
\sin^{2}(t/l)}\right\}^{2}\, ,
\label{6.3}
\end{eqnarray}
and
\begin{equation}
R_{abcd}\,R^{abcd}=\frac{({\cal D}-1)}{l^{4}}\,\left\{1 + \left[
\frac{\left[(M+1)^{2/({\cal D}-1)}-k\,r_{\Sigma}^{2}\right]\,
\cos^{2}(t/l) + k\,r_{\Sigma}^{2}}{\left[(M+1)^{2/({\cal
D}-1)}-k\,r_{\Sigma}^{2}\right]\,\sin^{2}(t/l)}\right]^{2}\right\}\, ,
\label{6.4}
\end{equation}
for the quadratic Ricci and Kretschmann scalars respectively. 
In this work we restrict the values of the quantity $k\,r_{\Sigma}^{2}$,
assuming $k\,r_{\Sigma}^{2}=0,\pm 1/2$.  These values have no special meaning, 
although for $k\,r_{\Sigma}^{2}$ positive and large enough there is no 
solution at all. Note also that the expression (\ref{5.14}) for the
mass is independent of the value chosen for $k\,r_{\Sigma}^{2}$.

Gravitational collapse occurs for $\pi/2 \leq 
t/l \leq \pi$.  The time $t/l=\pi/2$ marks the onset of
collapse. At this moment there are no singularities in spacetime, as
the curvature scalars (\ref{6.3})-(\ref{6.4}) and the density (\ref{6.2})
indicate. In fact, the singularity appears only at $t/l=\pi$, where
all these quantities blow up. 

To know whether a black hole has formed or not, one has to search for  
the appearance of an apparent horizon and an event horizon. 
The apparent horizon is defined to be the boundary 
of the region of trapped two-spheres in spacetime.
To find this boundary on the interior spacetime one looks for two spheres 
$Y\equiv a(t)r=$constant 
whose outward normals are null, i.e., 
$\nabla\,Y\,\cdot\,\nabla\,Y = 0\,$. 
Using metric (\ref{4.1}) this yields, 
\begin{equation}
\frac{d a(t)}{d t} = -\frac{\sqrt{1-k\,r^{2}}}{r}.
\label{6.6}
\end{equation}
Using (\ref{6.1}) in (\ref{6.6}) gives the evolution of the apparent 
horizon in comoving coordinates, 
\begin{equation}
\sqrt{\frac{(M+1)^{2/({\cal D}-1)} - k\,r_{\Sigma}^{2}}{1-k\,r_{\Sigma}^{2}\,
\left(r/r_{\Sigma}\right)^{2}}}\,\left(\frac{r}{r_{\Sigma}}\right) =
-\frac{1}{\cos\left(t/l \right)}\, ,
\label{6.7}
\end{equation}
Now, the apparent horizon first forms at the surface $r_\Sigma$. 
Then, for $r=r_\Sigma$, equation (\ref{6.7}) gives 
the time $t/l$ at which the apparent horizon first forms. 
On the other hand, one should also be able to find  
the formation time of the apparent horizon on the surface $\Sigma$ 
through an equation on $\Sigma$, 
equation (\ref{5.12}). Indeed, at the junction one has 
$R=a(t) r_\Sigma$. Then from junction condition (\ref{5.12}) and
equation (\ref{6.6}) we have that the apparent horizon first forms when 
\begin{equation}
\frac{R_{AH}}{l} = \sqrt{(M+1)^{2/({\cal D}-1)} - 1}\, .
	\label{6.8}
\end{equation}
Now, using (\ref{6.1}), the time of formation of 
the apparent horizon can be found through the equation 
\begin{equation}
\frac{R_{AH}}{l} = a(t_{AH})\,\frac{r_{\Sigma}}{l} =
\sqrt{(M+1)^{2/({\cal D}-1)}-k\,r_{\Sigma}^{2}}\,\sin\left(
\frac{t_{AH}}{l}\right)\, .
\label{6.9}
\end{equation}
Given a dimension ${\cal D}$ and an $M$ one can obtain 
$R_{AH}$ through equation (\ref{6.8}). Then equation (\ref{6.9}) 
gives implicitly $t_{AH}$, the time of the formation of the apparent 
horizon on the surface $\Sigma$ for a given $k$. For instance, for 
${\cal D}=3$, $M=0.25$ and $k\,r_{\Sigma}^{2}=0$ we find
$t_{AH}/l = 2.68$. Putting this value back in equation (\ref{6.7}) we 
verify that everything checks. 

The event horizon, being a null spherical surface, is determined through 
the null outgoing lines of the metric (\ref{4.1}), i.e., 
\begin{equation}
\frac{dt}{dr} = \frac{a(t)}{\sqrt{1-k\,r^{2}}}\, .
	\label{6.10}
\end{equation}
Equation (\ref{6.10}) can be put in the following integral form, 
\begin{equation}
\sqrt{(M+1)^{2/({\cal D}-1)} - k\,r_{\Sigma}^{2}}\,
\int_{0}^{r_{1}/r_{\Sigma}}\,\frac{d\left(r/r_{\Sigma}\right)}
{\sqrt{1-k\,r_{\Sigma}^{2}\left(r/r_{\Sigma}\right)^{2}}}=
\ln\left[\frac{\tan(x_{1})}{\tan(x_{0})}\right]\, ,
	\label{6.11}
\end{equation}
where $x \equiv (1/2)t/l$ . Now, the time $x_1$ is to be 
precisely equal to the formation time of the apparent horizon, since 
one expects that
in vacuum both horizons coincide \cite{hawkingellis}. One has then 
to integrate (\ref{6.11}) to find the time $x_0$ at which the event 
horizon first forms, at $r=0$. 
For instance, ${\cal D}=3$, $M=0.25$ and
$k\,r_{\Sigma}^{2}=-1/2$ 
we obtain $t_0 / l = 1.96$. A plot in comoving coordinates 
$(t/l,r/r_{\Sigma})$ shows the evolution of the apparent and event horizons
in Fig. 1. There we repeat the numerical calculations for the same value of
${\cal D}$ and $M$ but with $k\,r_{\Sigma}^{2}=0$ and $k\,r_{\Sigma}^{2}=
1/2$, as is shown in lines (b) and (c). In Fig. 2 we show
the formation of the apparent and event horizons for ${\cal D}= 25$, and $M=0.25$ and $k\,r_{\Sigma}^{2}=0$. Intermediate $\cal D$ dimensions 
have similar behavior.   
Making a matching to the vacuum exterior spacetime one finds the usual 
Penrose diagram for gravitational collapse and formation of a black 
hole in an anti-de Sitter background, see Fig. 3.

To study what happens to external observers we note that
a light signal emitted from the surface $r_{+}]_{\Sigma}$ 
at the exterior time $t_{+}$ obeys the
null condition
\begin{equation}
\frac{dr_{+}}{dt_{+}} = 1 - (M + 1)^{2/({\cal D}-1)} +
\left(\frac{r_{+}}{l}\right)^{2}\,,
\label{6.16}
\end{equation}
(see Eq. (\ref{3.4})). This light ray arrives at a point $r_{+}$ at time
$t_{+}$ given by
\begin{equation}
\frac{t_{+}}{l}=\frac{t_{+}]_{\Sigma}}{l} 
+ \frac{1}{2(M+1)^{2/({\cal D}-1)}-2}\,
\ln\left[\frac{(r_{+}/l) - [2(M+1)^{2/({\cal D}-1)}-2]}
{(r_{+}/l) + 
[2(M+1)^{2/({\cal D}-1)}-2]}\right]_{r_{+}]_{\Sigma}/l}^{r_{+}/l}\, .
\label{6.17}
\end{equation}
Thus $t_{+}/l \rightarrow \infty$ when $r_{+}]_{\Sigma}/l \rightarrow
\sqrt{(M+1)^{2/({\cal D}-1)}-1}$, so the collapse to the event horizon
appears to take an infinite amount of time to an exterior observer,
and the collapse to $r_{+}=0$ is
unobservable from the outside. Also, the redshift from the dust edge is
given by
\begin{equation}
z = \frac{dt_{+}}{dt}-1=\frac{1}{\sqrt{1-k\,r_{\Sigma}^{2}}+ \dot{R}} -1\, .
\label{6.18}
\end{equation}
When the dust edge crosses the event horizon we have
$\dot{R} = -\sqrt{1-k\,r_{\Sigma}^{2}}$, so $z \rightarrow \infty$. Thus
the collapsing dust will fade from sight, as the redshift of the light from its
surface diverges.

\section{Naked Singularities}

To study the presence of naked singularities, i.e., singularities not
hidden by an event horizon we analyse equations (\ref{3.4}),
(\ref{6.1})-(\ref{6.4}) and (\ref{5.14}). Naked singularities appear
only when $M<0$.  Although solutions with negative mass are usually
considered unphysical, they will be studied here because these
generalize the three-dimensional solutions found in
\cite{DJT,GAK,MannRoss}. In the model adopted here it is useful to
separate two distinct classes:

i) If $l$ remains finite (in which case $\Lambda \neq 0$), for any ${\cal D}
\geq 3$ the curvature scalars (\ref{6.3})-(\ref{6.4}) will blow up when
$t/l = \pi$, indicating the formation of a curvature naked singularity.

ii) If we take the limit $l \rightarrow \infty$ (in which case 
$\Lambda = 0$) we see from the exterior metric (\ref{3.4}) that the
event horizon is no longer present, and the collapse will form a naked
singularity.
Taking the limit on Eqs. (\ref{6.3})-(\ref{6.4}) we have
\begin{eqnarray}
&&R_{ab}\,R^{ab}=\frac{({\cal D}-1)\,({\cal D}-2)}{t^{4}}\,\left[
\frac{(M+1)^{2/({\cal D}-1)}}{(M+1)^{2/({\cal D}-1)}-k\,r_{\Sigma}^{2}}
\right]^{2}\, , \label{7.1} \\
&&R_{abcd}\,R^{abcd}=\frac{{\cal D}-1}{t^{4}}\,
\frac{(M+1)^{2/({\cal D}-1)}}{(M+1)^{2/({\cal D}-1)}-k\,r_{\Sigma}^{2}}\, .
\label{7.2}
\end{eqnarray}
For any ${\cal D} > 3$ both (\ref{7.1})-(\ref{7.2}) will vanish because 
from Eq. (\ref{5.14}) $M = -1 + {\cal O}(l^{-{\cal D}+3})$, so in the limit
we have $M = -1$. Also, from Eq. (\ref{6.1}) we have in the limit,  $a(t)=\sqrt{-k}\,t$, 
so that the only possible solution is when $k\,r_{\Sigma}^{2} < 0$.
Note also that $M=-1$ implies that the exterior metric (\ref{3.4}) is a
Minkowski one, although the interior density (\ref{6.2}) is non-zero
everywhere in the dust cloud. So at $t/l = \pi$ we will have $\rho \rightarrow
\infty$ in a flat Mikowski spacetime. This is analogous to a Newtonian
singularity.

For ${\cal D} =3$ we have $M=8\pi\rho_{0}(a_{0}r_{\Sigma})^{2}-1$ and
$a(t)=\sqrt{8\pi\rho_{0}a_{0}^{2}-k}\,t$, so that Eqs. (\ref{7.1})-(\ref{7.2})
will be finite but non-zero and the collapse will form a naked
conical singularity \cite{GAK,MannRoss}.

\section{Conclusions}

We have analysed gravitational collapse in Lovelock gravity for 
odd-dimensional spacetimes. We have showed that 
gravitational collapse of a regular initial non-rotating 
dust cloud proceeds, to form event and apparent horizons, and terminates 
at a spacelike curvature singularity.

\newpage

\centerline{\bf Figure Captions}
\vskip1cm

{\bf Figure 1.} 
Gravitational collapse in ${\cal D}=3$ dimensions in an asymptotically 
anti-de Sitter spacetime. The interior dust cloud in comoving coordinates 
$(t/l,r/r_{\Sigma})$ fills the whole diagram. The left side represents the
center of the 
cloud $r/r_{\Sigma}=0$, the right side the surface of the cloud
$r/r_{\Sigma}=1$. 
The evolution of the event horizon (dashed line) and apparent horizon 
(full line) are drawn. The singularity occurs at $t/l=0$. 
It was used $M=0.25$. The three different cases are 
(a) $k\,r_{\Sigma}^{2}=-1/2$, (b) $k\,r_{\Sigma}^{2}=0$, and 
(c) $k\,r_{\Sigma}^{2}=1/2$. 
\vskip .5cm

{\bf Figure 2.} Dimensionally continued Oppenheimer-Snyder collapse in 
${\cal D}=25$ dimensions in an asymptotically anti-de Sitter spacetime. 
It was used $M=0.25$ and $k\,r_{\Sigma}^{2}=0$. 
See subtitle of figure 1 for more detailed explanation. 
\vskip .5cm

{\bf Figure 3.} Penrose diagram for the collapse of a dust cloud in 
an asymptotically anti-de Sitter spacetime. Each point in the 
diagram represents a ${\cal D}-2$ sphere. (eh=event horizon, 
ah=apparent horizon).

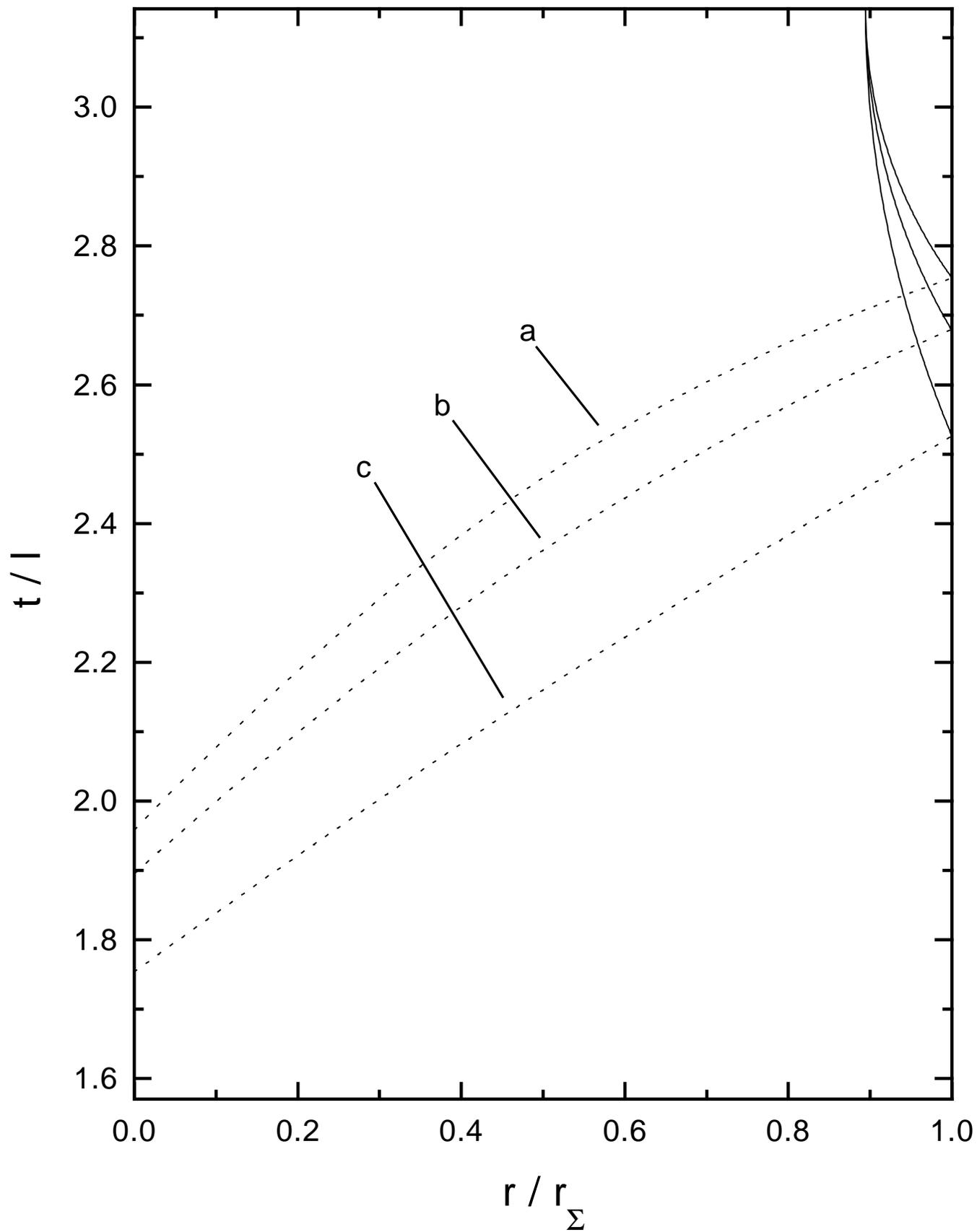
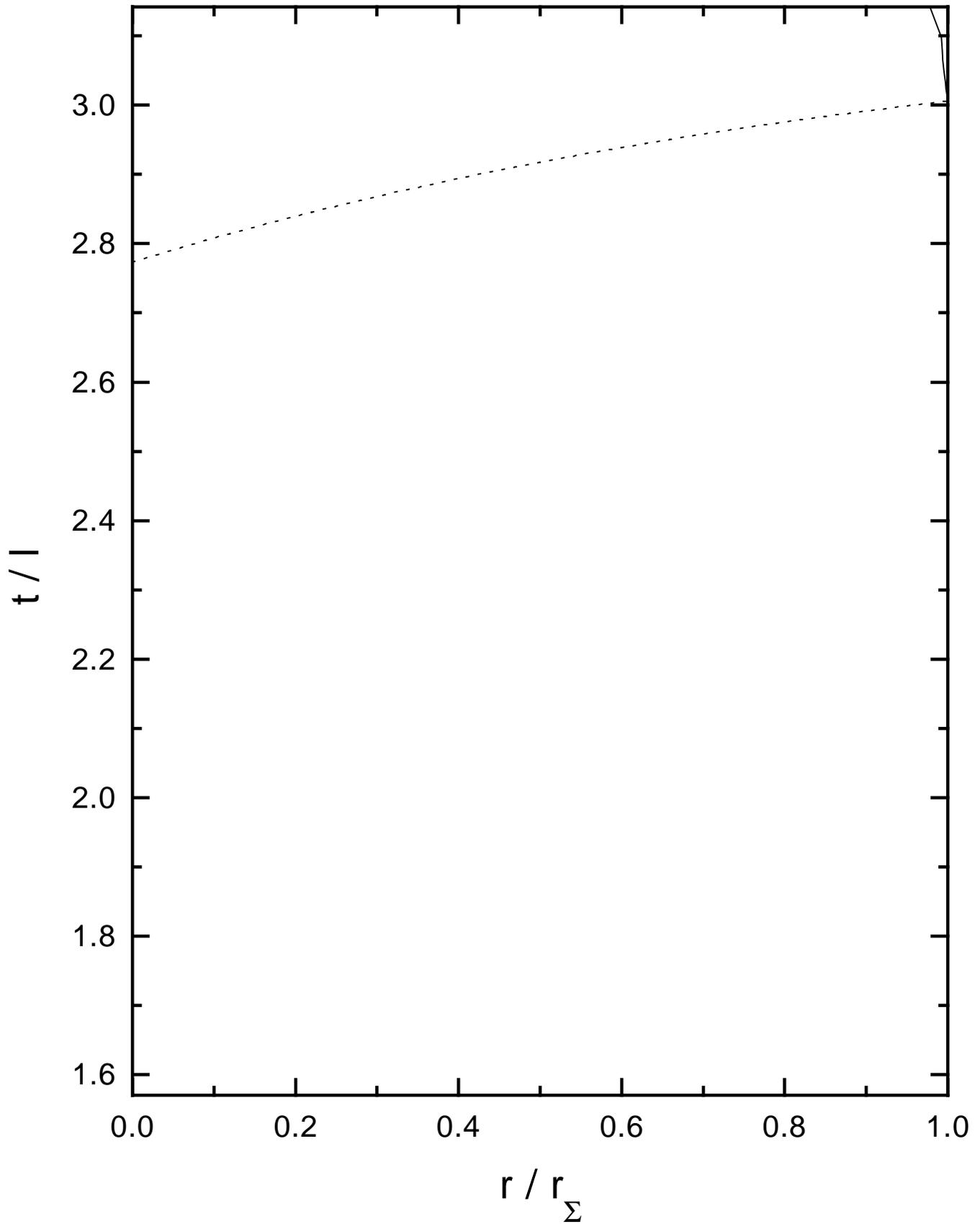
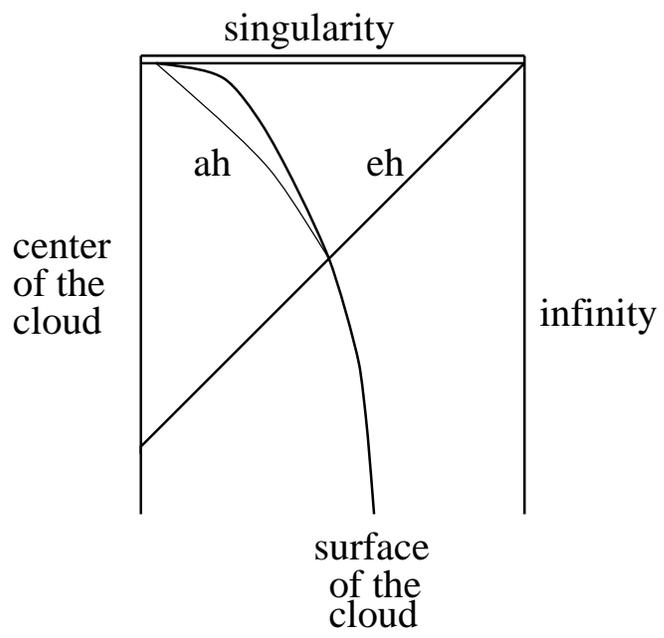
\end{document}